\documentstyle[preprint,floats,prd,aps]{revtex}
\input epsf.tex
\def\DESepsf(#1 width #2){\epsfxsize=#2 \epsfbox{#1}}
\begin{document}
\preprint{\vbox{\hbox{OITS-575, TECHNION-PH-95-12}}}
\draft
\title{ Energy Distribution of $\phi$ in Pure Penguin Induced $B$ Decays}
\author{N.G. Deshpande$^1$, G. Eilam$^2$, Xiao-Gang He$^1$,
and Josip Trampetic$^1$\footnote{On leave from Department of Theoretical
Physics,
R. Bo\v{s}kovi\'{c} Institute, Zagreb 41001, Croatia}}
\address{$^1$Institute of Theoretical Science\\
University of Oregon\\
Eugene, OR 97403-5203, USA\\
and\\
$^2$Physics Department,
Technion-Israel Institute of Technology\\
 32000, Haifa, Israel}

\date{May, 1995}
\maketitle
\begin{abstract}
We study the energy distribution of $\phi$ in pure penguin induced
$B\rightarrow
X_s \phi$ taking into account the fermi motion of b inside $B$ meson for
$b\rightarrow s\phi$ and also modification due to gluon bremsstrahlung
process $b\rightarrow s\phi g$. We
find that the contribution
to $B\rightarrow X_s\phi$ from $b\rightarrow s\phi g$ is less than 3\%. This
study provides a criterion for including most of the $\phi$'s produced in a
penguin process.

\end{abstract}
\newpage

Rare $B$ decays, particularly pure penguin decays, have been a subject of
considerable theoretical and experimental
interest recently\cite{1}. The photonic penguin processes have been observed
by
the CLEO collaboration\cite{2} in both the exclusive mode $B\rightarrow
K^*\gamma$
and in the inclusive mode $B\rightarrow X_s\gamma$. The Standard Model (SM) is
consistent with experimental data\cite{3}. A signature of pure penguin hadronic
processes are the exclusive modes $B\rightarrow K\phi$, $K^*\phi$\cite{gad}
etc. or the
inclusive  mode $B\rightarrow X_s \phi$\cite{4,5} and other modes resulting
from processes like $b\rightarrow \bar s ss$. The search for exclusive
processes
has not yet led to a definite observation. The inclusive mode with a larger
branching
ratio would be a complementary way of searching for penguin processes. At the
quark level $B\rightarrow X_s \phi$ results from $b\rightarrow s \phi$, just as
$B\rightarrow X_s \gamma$ results from $b\rightarrow s\gamma$. In both cases
the energy spectrum of $\phi$ or $\gamma$ are not monenergetic as a result of
two
effects. First, because $b$ quark is in the $B$ meson, its fermi momentum
smears the
energy spectrum of $\phi$ or $\gamma$. Therefore
the distribution of energy depends
to some extent on the choice of the wave function. Second effect arises
from
the process $b\rightarrow s g \gamma$ in the photonic case, and $b\rightarrow s
g \phi$ in the hadronic case. The photonic case has been discussed in a series
of papers by Ali and Greub\cite{ali}. They find that the dominant contribution
to the $\gamma$ spectrum comes from the wave function effect. We shall perform
a
similar calculation for the hadronic case. The wave function effect is treated
with a Monte-Carlo simulation of decays. The gluonic correction is carried out
in a simple effective Hamiltonian approximation. We find that
the second effect is
negligible in our case.

The QCD corrected $H_{\Delta B = 1}$ relevant to us
can be written as follows\cite{6}:
\begin{eqnarray}
H_{\Delta B=1} = {G_F\over \sqrt{2}}[V_{ub}V^*_{us}(c_1O^u_1 + c_2 O^u_2)
+V_{cb}V^*_{cs}(c_1O^c_1+c_2O^c_2) - V_{tb}V^*_{ts}\sum c_iO_i] +H.C.\;,
\end{eqnarray}
where the Wilson coefficients (WCs) $c_i$ are defined at the scale of $\mu
\approx
m_b$; and $O_i$ are defined as
\begin{eqnarray}
O^q_1 = \bar s_\alpha \gamma_\mu(1-\gamma_5)q_\beta\bar
q_\beta\gamma^\mu(1-\gamma_5)b_\alpha\;&,&\;\;\;\;
O^q_2 = \bar s \gamma_\mu(1-\gamma_5)q\bar
q\gamma^\mu(1-\gamma_5)b\;,\nonumber\\
O_{3,5} = \bar s \gamma_\mu(1-\gamma_5)b \sum_{q'}
\bar q' \gamma_\mu(1\mp\gamma_5) q'\;&,&\;\;\;\;
Q_{4,6} = \bar s_\alpha \gamma_\mu(1-\gamma_5)b_\beta \sum_{q'}
\bar q'_\beta \gamma_\mu(1\mp\gamma_5) q'_\alpha\;,\nonumber\\
O_{7,9} ={3\over 2}\bar s \gamma_\mu(1-\gamma_5)b \sum_{q'} e_{q'}\bar q'
\gamma^\mu(1\pm\gamma_5)q'\;&,&\;\;
Q_{8,10} = {3\over 2}\bar s_\alpha \gamma_\mu(1-\gamma_5)b_\beta \sum_{q'}
e_{q'}\bar q'_\beta \gamma_\mu(1\pm\gamma_5) q'_\alpha\;.
\end{eqnarray}

The Wilson coefficients at $\mu = m_b$ at the next-to-leading order have been
evaluated in Refs.\cite{5,6,7}. For $m_t = 176$ GeV and $\alpha_s(m_Z) =
0.117$, we
find
\begin{eqnarray}
c_1 &=& -0.307\;, c_2 = 1.147\;, c_3 = 0.017\;, c_4 = -0.037\;, c_5 = 0.010
\nonumber\\
c_6 &=& -0.045\;, c_7 = 1.2\times 10^{-5}\;, c_8 = 3.8\times 10^{-4}\;,
c_9 = -0.010\;, c_{10} = 2.1\times 10^{-3}\;.
\end{eqnarray}

We shall consider the effect due to fermi momentum in $b\rightarrow s\phi$
process, which we assume has the same $\phi$ energy distribution as
$B\rightarrow X_s \phi$, in  section (a) below. In section (b) we shall
consider the effect due
to $b\rightarrow s\phi g$ process.\\

\noindent
{\large{\bf a. $b\rightarrow s\phi$}}

Using $H_{\Delta B = 1}$ in Eq.(1), we obtain the
decay amplitude for $B\rightarrow X_s \phi$
\begin{eqnarray}
A(b\rightarrow s \phi)
= - a \bar s \gamma_\mu (1-\gamma_5)b\phi^\mu\;,
\end{eqnarray}
where $\epsilon^\mu$ is the polarization
of the $\phi$ particle; $a = (g_\phi
G_FV_{tb}V_{ts}^*/\sqrt{2})[c_3+c_4+c_5+\xi
(c_3+c_4+c_6) - (c_7+c_9+c_{10}
+\xi(c_8 +c_9+c_{10}))/2]$ with $\xi=1/N_c$,
and $N_c$ is the number of colors.
The coupling constant $g_\phi$ is defined
by $<\phi|\bar s\gamma^\mu s|0> = i g_\phi \epsilon^\mu$. From the experimental
value for $Br(\phi \rightarrow e^+e^-)$\cite{8}, we obtain
$g^2_\phi = 0.0586\; GeV^4$. The branching ratio for $b\rightarrow s\phi$ is
predicted to be $1.7\times
10^{-4}$\cite{5} for $\alpha_s(m_Z) = 0.117$.

The decay rate is given by
\begin{eqnarray}
\Gamma(b \rightarrow s\phi) = {|a|^2m_b^3 \over 8\pi m_\phi^2}
\lambda_{s\phi}^{3/2}[1 + {3\over \lambda_{s\phi}}
{m_\phi^2\over m_b^2}(1-{m_\phi^2\over m_b^2} +{m_s^2\over m_b^2})]\;,
\end{eqnarray}
where $\lambda_{ij} = (1-m_j^2/m_b^2 -m_i^2/m_b^2)^2 - 4m_i^2m_j^2/m_b^4$.

To study the energy distribution of $\phi$,
we adopt the model in Ref.\cite{9} in which the b quark is not at rest inside
$B$
but with a fermi momentum $p_b$ according to a Gaussian distribution,
\begin{eqnarray}
\Phi (\vec p_b) = {4\over \sqrt{\pi}p_f^3} e^{-\vec p_b^2/p_f^2}\;,
\end{eqnarray}
where the parameter $p_f$ is determined from experimental data to be between
0.21 to 0.39 GeV\cite{10}.
The b quark mass expressed in terms of the $B$ meson mass $m_B$ and the
spectator quark mass $m_q$ in the rest frame of $B$, is given by
\begin{eqnarray}
m_b^2 = m_B^2 +m_q^2 - 2 m_B \sqrt{ \vec p_b^2 +m_q^2}\;.
\end{eqnarray}
In the rest frame of the $B$, the $b\rightarrow s\phi$ decay width
$\Gamma(m_b)$ is
given by $(m_b/E_b)\Gamma(b\rightarrow s \phi)$.
Its contribution to the decay width $\Gamma_{s\phi}$ for $B\rightarrow X_s\phi$
is
averaged over all allowed momenta
$\vec p_b$. We have
\begin{eqnarray}
\Gamma_{s\phi}(B \rightarrow  X_s \phi)
&=& {\large \int}^{P_{max}}_0 \Phi(\vec p_b) \vec p_b^2 \Gamma(m_b) d |\vec
p_b|\;,\nonumber\\
P_{max} &=& \sqrt{{(m_B^2+m_q^2 -(m_\phi +m_s)^2)^2\over 4 m_B^2} - m_q^2}\;.
\end{eqnarray}
Due to the finite momentum distribution, the energy of $\phi$ from b quark
decay is
no longer monoenergetic, instead there will be a distribution. The $\phi$
energy spectrum generated from a Monte-Carlo simulation of the decay is shown
in
Figure 1. In the figures we have used the constituent mass
of 0.3 GeV and 0.5 GeV for the spectator quark and the s-quark, respectively.
{}From Fig.1, we see indeed that there is a spread in the $\phi$ energy with
the
maximum located at about 2.55 GeV.\\

\noindent
{\large{\bf b. $b\rightarrow s\phi g$}}

The quark level effective Hamiltonian responsible for $b \rightarrow s \phi g$
is
complicated. We use the simplified effective Hamiltonian in Eq.(4) to obtain
the $\phi$ energy distribution for the process $b\rightarrow s\phi g$ by
attaching
a gluon on either the initial b or the final s quarks.  This is
expected to be a
good approximation because the dominant effect comes from the
bremsstrahlung of gluon emission from the external light quark. We obtain
\begin{eqnarray}
A (b\rightarrow s\phi g) &=& a g_s ({\bar s \gamma_\mu (2p_b^\nu - \not p_g
\gamma^\nu) T^a (1-\gamma_5) b \over (p_b-p_g)^2-m_b^2}\nonumber\\
&+& {\bar s (2p_s^\nu +\gamma^\nu \not p_g)\gamma_\mu T^a(1-\gamma_5) b
\over (p_s+p_g)^2-m_s^2}) G^a_\nu \phi^\mu\;,
\end{eqnarray}
where $p_b\;,p_s\;$, and $p_g$ are the b-quark, s-quark and gluon
momenta, respectively. From above we have the following
$\phi$ energy spectrum
\begin{eqnarray}
&&{d\Gamma(b\rightarrow s\phi g)\over d E_\phi} =
{|a|^2 \alpha_s\over 32 \pi^2 m_b^2}{N_c^2-1\over N_c^2} {\large
\int}^{t_{max}}_{t_{min}} d t
[{1\over 1+Y}(4 + (1+Y)^2 +(1-Y)^2{1+\mu_s^2\over 2\mu_\phi^2})\nonumber\\
 &&+{2\over X^2(1+Y)}(1-2\mu_\phi^2-X-Y+XY - 2\mu_s^2{1-Y\over 1+Y})
({(1-\mu_s^2)^2\over \mu_\phi^2} - 2\mu^2_\phi + 1+\mu^2_s)]\;,
\end{eqnarray}
where
\begin{eqnarray}
&X &= {s+t - m_s^2-m_\phi^2\over m_b^2}\;,\;Y = {s+t - m_s^2-m_\phi^2 \over s-t
-m_s^2 +m_\phi^2}\;,\nonumber\\
&s &= m_b^2 +m_\phi^2 - 2m_b E_\phi\;, \; \mu_{s,\phi}= {m_{s,\phi} \over
m_b}\;,\nonumber\\
&t&_{max, min} = {(s-m_s^2)\over 2s}((m_b^2-s-m_\phi^2) \pm
\sqrt{(m_b^2-s-m_\phi^2)^2 - 4 m_\phi^2 s}) + m_\phi^2\;.
\end{eqnarray}
The energy distribution in Eq.(10)
has the well-known infrared divergence due to the zero
mass of the gluon. To regulate the infrared divergence, we assign an effective
gluon mass of about $2m_\pi$ which represents the lowest invariant mass of the
gluon. The $\phi$ energy distribution for $b \rightarrow s\phi g$ is shown in
Figure 2. Here we
have neglected the effect due to non-zero $\vec p_b$ discussed in the previous
section which is small and approximated the $b\rightarrow s\phi g$
contribution to $B\rightarrow X_s \phi$ by Eq.(10). We find that
the effect of $b\rightarrow s\phi g$ on $B\rightarrow X_s\phi$ is small
because $BR(b\rightarrow s\phi g)/BR(b\rightarrow s\phi)$ is only about 3\%.
The total energy distribution is shown in Figure 3.

The spectrum of $\phi$ should approximate the spectrum that arises from the
decays $B\rightarrow K\pi \phi$, $K\pi\pi\phi$, etc. Of course the
monoenergetic $\phi$'s that arise from two body modes like $B\rightarrow K\phi$
or $K^*\phi$ are included in an average sense. The specific two body modes are
not expected to be more than 10\% of the inclusive $X_s\phi$
production\cite{5}. In this paper we have not discussed the $\phi$ energy
spectrum from the decay of the dominant non-penguin processes which are
expected to have a much softer spectrum since they always arise from decay of
charmed states.
We assume that this experimentally well known contribution has been
substracted
in the region of interest. If in addition to the selection criterion on $\phi$
discussed in this letter, $X_s$ will be experimentally shown to include an odd
number of kaons, then the penguin process will be even more
enhanced\cite{gad}.

\acknowledgments
NGD and XGH thank Drs. Z. Chao, R. Frey and D. Strom for helps in computer
simulations.  NGD, XGH and JT were supported in part by the Department of
Energy
Grant No. DE-FG06-85ER40224. GE was supported in part by GIF-The
German-Israeli Foundation for Scientific Research and Development and by
the New York Metropolitan Research Fund.

\begin{figure}[htb]
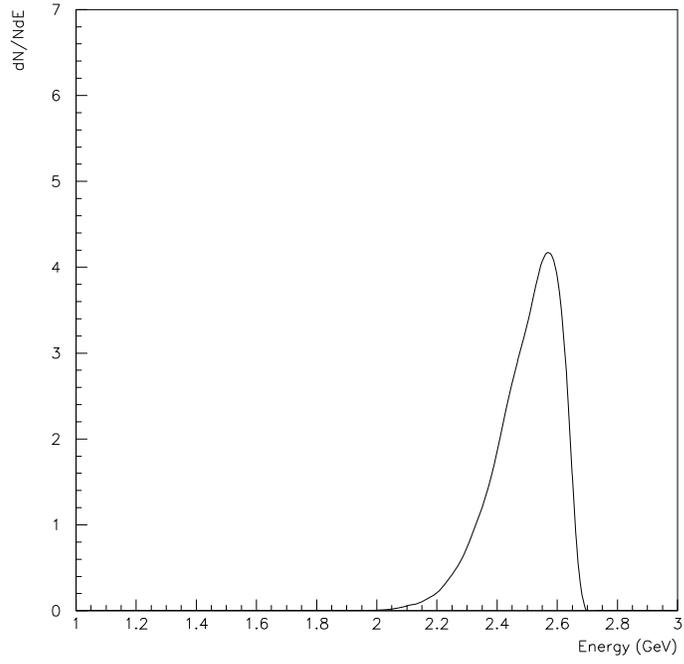

\centerline{ \DESepsf(bophi1.epsf width 10 cm) }
\smallskip
\caption{$E_\phi$ distribution for $b\rightarrow s \phi$ for $p_f$ = 0.3 GeV. }
\end{figure}

\begin{figure}[htb]
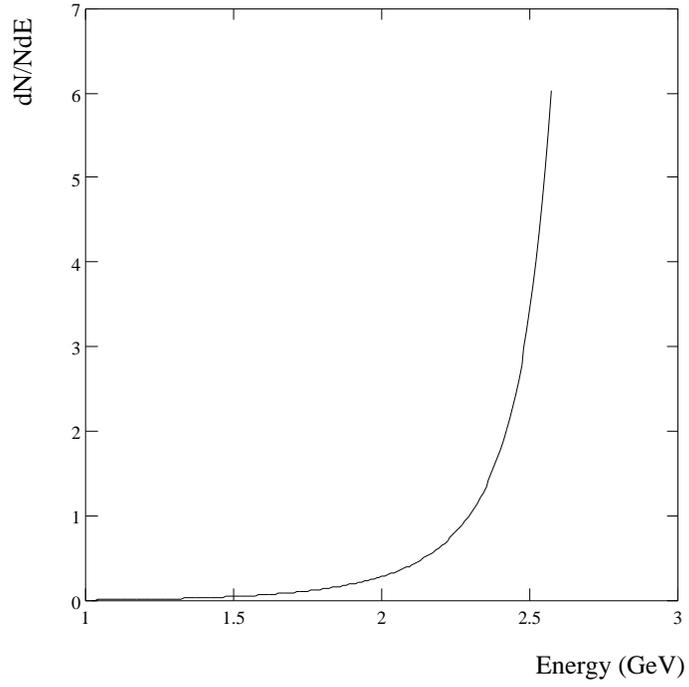

\centerline{ \DESepsf(bophi2.epsf width 8.75 cm) }
\smallskip
\caption{$E_\phi$ distribution for $b\rightarrow s \phi g$.  }
\end{figure}

\begin{figure}[htb]
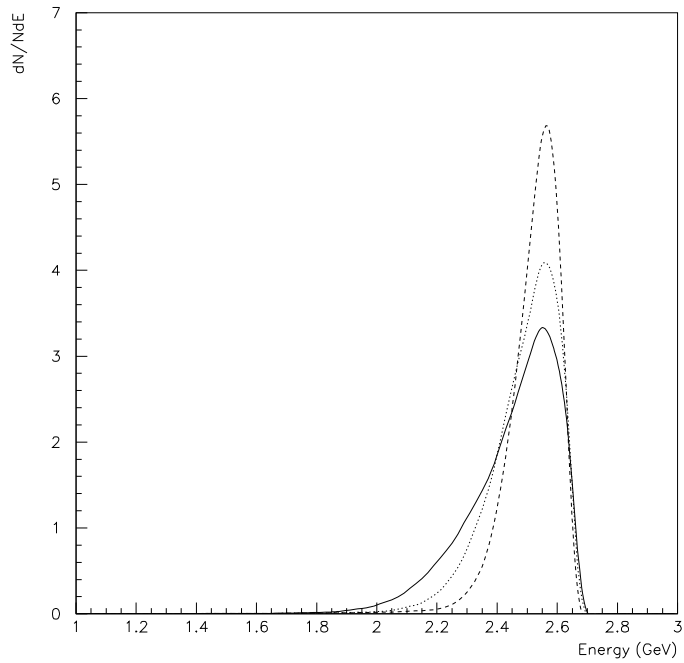

\centerline{ \DESepsf(bophi3.epsf width 10 cm) }
\smallskip
\caption{$E_\phi$ distribution for $b\rightarrow s \phi + s \phi g$. The solid,
doted and
dashed lines are for $p_f$= 0.39 GeV, 0.30 GeV and 0.21 GeV, respectively.  }
\end{figure}

\end{document}